\documentclass[a4paper,fleqn,usenatbib]{mnras}

\usepackage{newtxtext,newtxmath}

\usepackage[T1]{fontenc}
\usepackage{ae,aecompl}

\usepackage{graphicx}	
\usepackage{amsmath}	
\usepackage{amssymb}	

\def\chisqr{\hbox{$\chi^2_{\rm r}$}}
\def\msun{\hbox{${\rm M}_{\odot}$}}

\def\mspy{\hbox{${\rm M}_{\odot}$\,yr$^{-1}$}}
\def\rsun{\hbox{${\rm R}_{\odot}$}}
\def\lsun{\hbox{${\rm L}_{\odot}$}}
\def\rcor{\hbox{$r_{\rm cor}$}}
\def\rmag{\hbox{$r_{\rm mag}$}}
\def\mstar{\hbox{$M_{\star}$}}
\def\rstar{\hbox{$R_{\star}$}}
\def\lstar{\hbox{$L_{\star}$}}
\def\teff{\hbox{$T_{\rm eff}$}}
\def\logg{\hbox{$\log g$}}
\def\sn{\hbox{S/N}}
\def\vrad{\hbox{$v_{\rm rad}$}}
\def\vesc{\hbox{$v_{\rm esc}$}}

\def\kms{\hbox{km\,s$^{-1}$}}
\def\vsini{\hbox{$v \sin i$}}

\def\V{\hbox{${\rm V}$}}
\def\BmV{\hbox{${\rm B-V}$}}

\def\AV{\hbox{$A_{\rm V}$}}

\def\Bl{\hbox{$B_{\rm \ell}$}}

\def\degr{\hbox{$^\circ$}}

\def\Mdot{\hbox{$\dot{M}$}}

\def\Prot{\hbox{$P_{\rm rot}$}}

\newcommand{\caii}{\hbox{Ca$\;${\sc ii}}}

\newcommand{\oxi}{\hbox{O$\;${\sc i}}}

\newcommand{\hei}{\hbox{He$\;${\sc i}}}
\newcommand{\hal}{\hbox{H${\alpha}$}}
\newcommand{\hbe}{\hbox{H${\beta}$}}

\title[The magnetic field and accretion regime of CI~Tau]{The magnetic field and accretion regime of CI~Tau} 
\author[J.-F.~Donati et al.]{J.-F.~Donati$^{1}$\thanks{E-mail: jean-francois.donati@irap.omp.eu},
           J.~Bouvier$^{2}$, S.H.~Alencar$^{3}$, C.~Moutou$^1$, L.~Malo$^4$, M.~Takami$^{5}$, 
\newauthor F.~M\'enard$^{2}$, C.~Dougados$^{2}$, G.A.~Hussain$^7$ and the MaTYSSE collaboration 
\vspace{1mm}
\\ 
$^1$ Univ.\ de Toulouse, CNRS, IRAP, 14 avenue Belin, 31400 Toulouse, France \\ 
$^2$ Univ.\ Grenoble Alpes, CNRS, IPAG, 38000 Grenoble, France \\ 
$^3$ Departamento de F\`{\i}sica -- ICEx -- UFMG, Av. Ant\^onio Carlos, 6627, 30270-901 Belo Horizonte, MG, Brazil \\
$^4$ D\'epartement de physique, Universit\'e de Montr\'eal, C.P.~6128, Succursale Centre-Ville, Montr\'eal, QC, Canada  H3C 3J7 \\
$^5$ Institute of Astronomy and Astrophysics, Academia Sinica, PO Box 23-141, 106, Taipei, Taiwan \\
$^6$ ESO, Karl-Schwarzschild-Str.\ 2, D-85748 Garching, Germany 
}

\date{Submitted 2019 xxx -- Accepted 2019 xxx } 

\pubyear{2019}

\begin{document}

\label{firstpage}
\pagerange{\pageref{firstpage}--\pageref{lastpage}}
\maketitle

\begin{abstract}
This paper exploits spectropolarimetric data of the classical T~Tauri star CI~Tau collected with ESPaDOnS at the 
Canada-France-Hawaii Telescope, with the aims of detecting and characterizing the large-scale magnetic field that the 
star hosts, and of investigating how the star interacts with the inner regions of its accretion disc through this field.  
Our data unambiguously show that CI~Tau has a rotation period of 9.0~d, and that it hosts a strong, mainly poloidal 
large-scale field.  Accretion at the surface of the star concentrates within a bright high-latitude chromospheric region 
that spatially overlaps with a large dark photospheric spot, in which the radial magnetic field reaches $-3.7$~kG.  
\color{black}
With a polar strength of $-1.7$~kG, the dipole component of the large-scale field is able to evacuate the central 
regions of the disc up to about 50\%\ of the co-rotation radius (at which the Keplerian orbital period equals the stellar 
rotation period) throughout our observations, during which the average accretion rate was found to be unusually high.  
We speculate that the magnetic field of CI~Tau is strong enough to sustain most of the time a magnetospheric gap extending 
to at least 70\%\ of the co-rotation radius, which would explain why the rotation period of CI~Tau is as long as 9~d.  
\color{black}
Our results also imply that the 9-d radial velocity (RV) modulation that CI~Tau exhibits is attributable to stellar 
activity, and thus that the existence of the candidate close-in massive planet CI~Tau~b to which these RV fluctuations 
were first attributed needs to be reassessed with new evidence.  
\end{abstract}

\begin{keywords}
stars: magnetic fields --
stars: formation --
stars: imaging --
stars: individual:  CI~Tau  --
techniques: polarimetric
\end{keywords}



\section{Introduction}
\label{sec:int}

Young protostars and pre-main-sequence (PMS) stars are ideal laboratories to assess theories of star / planet formation.  This is 
especially true for investigating the key role that magnetic fields play in this process, e.g., when most of the initial angular 
momentum and magnetic flux inherited from the parent molecular cloud is removed from the central core \citep[e.g.,][]{Hennebelle08b, 
Vaytet18}.  At a later stage, the large-scale field of the star is found to be strong enough to evacuate the central regions of the 
accretion disc, forcing the star to co-rotate with the inner disc \citep[e.g.,][]{Bouvier14, Hartmann16}, whereas disc fields 
are potentially able to affect how planets, including massive giants, form and migrate within the disc \citep[e.g.,][]{Lin96, Muto08}.  

Both classical T~Tauri stars (cTTSs), still accreting from their surrounding accretion discs, and weak-line T~Tauri stars (wTTSs), no 
longer accreting from their depleted inner disc or having exhausted their disc entirely, are key targets for such studies.  By unveiling 
their large-scale magnetic topologies using tomographic techniques such as Zeeman-Doppler Imaging \citep[ZDI,][]{Semel89, Brown91, Donati97c,
Donati06b} applied to phase-resolved spectropolarimetric data sets, one can quantitatively investigate how cTTSs interact and exchange 
angular momentum with their discs \citep[e.g.,][]{Donati11};  by modelling their magnetic activity and in particular their surface brightness 
inhomogeneities, one can also reveal the potential presence of close-in giant planets orbiting wTTSs \citep[e.g.,][]{David16, Donati17, Yu17}.  
Last but not least, one can study how dynamo processes operate within the convective interior / envelope of TTSs and generate their large-scale 
fields, to be compared with those of mature main-sequence low-mass stars \citep[e.g.,][]{Morin08b, Donati09, Morin10, Donati11, Gregory12, Folsom16}.  

Among the handful of cTTSs identified for observations within the MaPP and MaTYSSE Large Programmes carried out at Canada-France-Hawaii 
Telescope (CFHT) with the ESPaDOnS spectropolarimeter, CI~Tau, reported to potentially host a close-in giant planet \citep{JohnsKrull16, 
Biddle18, Flagg19}, is thus especially interesting for tomographic studies based on spectropolarimetric monitoring.  In this paper, we present 
ESPaDOnS observations of CI~Tau collected from mid December 2016 to mid February 2017;  following a short review of the evolutionary status of 
this cTTS, we analyze our new data in terms of the large-scale magnetic topologies that the newborn star hosts, and of the accretion patterns that 
the star-disc interactions trigger.  We finally conclude our study with implications for our understanding of magnetospheric accretion in cTTSs, 
and for the candidate close-in giant planet CI~Tau~b.

\section{Evolutionary status of CI~Tau}
\label{sec:evo}

To put our study on a firm footing and make it consistent with previous work, we start by revising the evolutionary status of CI~Tau.  

We apply to our best spectra of CI~Tau (see Sec.~\ref{sec:obs}) the spectral classification tool we developed and used previously  
\citep[e.g.,][]{Donati12}; we derive a photospheric temperature of $\teff=4200\pm50$~K and a logarithmic surface gravity of $\logg=3.6\pm0.2$ 
(in cgs units).  We note that our temperature estimate is larger than the value often quoted in the literature \citep[of 4060~K][]{Kenyon95} 
but in good agreement with independent measurements from high-resolution spectroscopy \citep{Schiavon95}, or more recent values from 
low-resolution spectrophotometry \citep{Herczeg14}.  Given the published \BmV\ color index of CI~Tau \citep[1.37,][]{Grankin07} and the one 
we expect for a young star of the same temperature \citep[$1.16\pm0.02$,][]{Pecaut13}, we derive that the visual extinction the star suffers 
is $\AV=0.65\pm0.20$\footnote{In doing so, we assume that veiling (see Sec.~\ref{sec:obs}) and cool surface spots (see Sec.~\ref{sec:tom}), 
having opposite effects on photometric colors, more or less cancel out within the quoted error bar on \AV.}.  The corresponding bolometric 
correction is equal to $-0.89\pm0.02$ \citep{Pecaut13}. 

From the minimum \V\ magnitude (maximum brightness) of CI~Tau \citep[$\V=12.28\pm0.10$,][]{Grankin07} and assuming that CI~Tau always features 
cool surface spots making it dimmer by at least $0.25\pm0.10$~mag in \V\footnote{Here we refer not only to large-scale spots like those 
reconstructed in Sec.~\ref{sec:tom}, but also to small-scale ones evenly distributed over the star and generating little to no rotational 
modulation as in, e.g., \citet{Gully17}.  The chosen value of $0.25\pm0.10$ corresponds to the residual magnitude increase that is not 
compensated by minimum veiling at maximum brightness (see Sec.~\ref{sec:tom}), implying that the dimming from dark spots only is in fact 
larger.  \color{black} The average \V\ magnitude of CI~Tau \citep[$\V=13.11\pm0.10$,]{Grankin07} indicates that about two thirds of the stellar 
surface are usually covered with spots.}\color{black}, 
we derive a bolometric magnitude of $4.49\pm0.25$ using the accurate distance estimate from Gaia \citep[$158.7\pm1.2$~pc,][]{Gaia18}.  
The corresponding logarithmic luminosity relative to the Sun ($\log(\lstar/\lsun)=0.1\pm0.1$) translates into a mass and radius of CI~Tau equal 
to $\mstar=0.9\pm0.1$~\msun\ and $\rstar=2.0\pm0.3$~\rsun\ respectively when using the PMS evolutionary models of \citet[][assuming solar 
metallicity and including convective overshooting]{Siess00};  in this context, CI~Tau is a $2\pm1$~Myr star that has not yet started to 
develop an inner radiative core.  Our mass estimate is in very good agreement with the dynamic value derived from radio interferometry 
\citep[equal to $0.90\pm0.02$~\msun,][which we use hereafter]{Simon17, Simon19};  our luminosity estimate is larger than the literature 
values \citep{Andrews13,Herczeg14}, reflecting both the revised distance and our choice to compensate for the contribution of spots at the 
stellar surface.  

The rotation period we derive for CI~Tau ($\Prot=9.00\pm0.05$~d, see Sec.~\ref{sec:obs}) and the line-of-sight-projected equatorial rotation 
velocity we measure ($\vsini=9.5\pm0.5$~\kms, see Sec.~\ref{sec:tom}, with $i$ noting the angle between the line of sight and the rotation 
axis of the star) imply that $\rstar \sin i=1.69\pm0.10$~\rsun\ and thus that $i$$\simeq$55\degr, with potential values ranging from 45\degr\ 
up to 90\degr.  This is consistent with the tilt of the disc rotation axis to the line of sight, reported to be in the 
range 46\degr\ to 54\degr\ \citep{Guilloteau14, Clarke18}.  This agreement further supports our luminosity estimate and the associated radius 
value, and may even suggest that both are still slightly underestimated;  given the reported values of \vsini\ and \Prot, a lower luminosity and 
a smaller radius would indeed imply that the inclination of the star significantly differs from that of the disc.  
\color{black}This places the co-rotation radius of CI~Tau (i.e., the radius at which the Keplerian orbital period equals the stellar rotation) 
at a distance of $\rcor=0.082\pm0.001$~au or $8.8\pm1.3$~\rstar\ from the centre of the star, thereby setting the co-rotation velocity 
at $99\pm1$~\kms\ and its line-of-sight projection at $76\pm5$~\kms\ for a disc inclination of $50\pm4$~\degr\ \citep{Guilloteau14, Clarke18}, 
consistent with the RV semi-amplitude of the CO signature recently reported \citep[equal to 77~\kms,][]{Flagg19}.  \color{black}

The fundamental parameters of CI~Tau used in our study are summarized in Table~\ref{tab:par}.  

\begin{table}
\caption[]{Summary of the fundamental parameters of CI~Tau used in our study (all estimates are from our work except where noted)} 
\hspace{-6mm}
\begin{tabular}{ccc}
\hline
\teff\ (K)           & $4200\pm50$   & \\
$\log(\lstar/\lsun)$ & $0.1\pm0.1$   & \\ 
\mstar\ (\msun)      & $0.9\pm0.1$   & \citet{Siess00} \\
                     & $0.90\pm0.02$   & \citet{Simon19} \\
\rstar\ (\rsun)      & $2.0\pm0.3$   & \citet{Siess00} \\
age (Myr)            & $2.0\pm1.0$   & \citet{Siess00} \\
distance (pc)        & $158.7\pm1.2$ & \citet{Gaia18} \\
\Prot\ (d)           & $9.00\pm0.05$ & \\ 
$i$ (\degr)          & $55^{+35}_{-10}$ & \\ 
\color{black} $i_{\rm disc}$ (\degr) & \color{black} $50\pm4$ & \color{black} \citet{Guilloteau14}  \\ 
\vsini\ (\kms)       & $9.5\pm0.5$   & \\ 
\vrad\  (\kms)       & $16.8\pm0.2$  & \\
\color{black} \vesc\  (\kms)       & \color{black} $415\pm30$  & \\
\color{black} \rcor\ (au)      & \color{black} $0.082\pm0.001$ & \color{black} $8.8\pm1.3$~\rstar\\ 
\color{black} $\log \Mdot$ (\mspy) & \color{black} $-7.6\pm0.3$  & \color{black} from \hal, \hbe\ and \hei\ $D_3$ \\
\hline
\end{tabular}
\label{tab:par}
\end{table}

\begin{table*}
\caption[]{Journal of ESPaDOnS observations of CI~Tau from December 2016 to February 2017.  
All observations consist of sequences of 4 subexposures, each lasting either 1200~s (first 8 spectra) or 1000~s (last 10 spectra).  
Columns respectively list, for each observation, the UT date, time, Barycentric Julian Date (BJD), peak signal to noise ratio \sn\ 
(per 2.6~\kms\ velocity bin), rms noise level in Stokes $V$ LSD profiles, rotation cycle $c$ computed using ephemeris BJD~(d)~=~$2457736.7+9.0 c$, 
longitudinal fields \Bl\ measured from both LSD profiles and \hei\ $D_3$ narrow emission component (NC), the RVs and bisector spans (BSs) 
of LSD profiles, and the RVs of the \hei\ $D_3$ NC. }
\hspace{-6mm}
\begin{tabular}{ccccccccccc}
\hline
Date   &    UT      & BJD          & \sn\ & $\sigma_{\rm LSD}$ & $c$ & $\Bl_{\rm LSD}$ & $\Bl_{\rm He}$ & RV$_{\rm LSD}$ & BS$_{\rm LSD}$ & RV$_{\rm He}$ \\
       & (hh:mm:ss) & (2,457,736+) &      &   (0.01\%)         &     &      (G)        &   (kG)         &      (\kms)    &       (\kms)   &    (\kms)     \\
\hline
2016 Dec 14 & 07:01:06 &  0.79801 & 160 & 3.0 & 0.011 & $95\pm11$  & $-1.15\pm0.49$ & $16.7\pm0.3$ & $0.0\pm0.3$  & $22.1\pm0.3$ \\
2016 Dec 15 & 07:08:38 &  1.80322 & 160 & 3.0 & 0.123 & $70\pm10$  & $-1.70\pm0.48$ & $17.4\pm0.3$ & $-0.3\pm0.3$ & $21.4\pm0.3$ \\
2016 Dec 21 & 08:12:55 &  7.84769 & 170 & 3.0 & 0.794 & $48\pm11$  & $-1.31\pm0.50$ & $16.8\pm0.3$ & $-0.3\pm0.3$ & $27.6\pm0.3$ \\
2016 Dec 22 & 08:30:34 &  8.85991 & 180 & 2.6 & 0.907 & $15\pm11$  & $-2.46\pm0.74$ & $17.3\pm0.3$ & $-0.7\pm0.3$ & $24.4\pm0.3$ \\
2017 Jan 08 & 08:53:59 & 25.87534 & 180 & 2.7 & 2.797 & $-21\pm10$ & $-0.55\pm0.66$ & $17.4\pm0.3$ & $-0.5\pm0.3$ & $23.2\pm0.3$ \\
2017 Jan 09 & 09:50:38 & 26.91461 & 180 & 2.6 & 2.913 & $-3\pm10$  & $-0.98\pm0.61$ & $16.9\pm0.3$ & $-0.0\pm0.3$ & $23.6\pm0.3$ \\
2017 Jan 11 & 06:10:15 & 28.76144 & 150 & 3.3 & 3.118 & $95\pm12$  & $-3.49\pm0.52$ & $17.7\pm0.3$ & $-0.3\pm0.3$ & $22.8\pm0.3$ \\
2017 Jan 13 & 11:37:52 & 30.98881 & 160 & 3.0 & 3.365 & $-25\pm13$ & $-1.27\pm0.38$ & $17.7\pm0.3$ & $-0.7\pm0.3$ & $22.8\pm0.3$ \\
\hline
2017 Jan 15 & 10:47:14 & 32.95351 & 140 & 3.6 & 3.584 & $60\pm16$  & $0.05\pm0.46$  & $14.8\pm0.3$ & $1.2\pm0.3$  & $27.6\pm0.3$ \\
2017 Jan 16 & 09:53:01 & 33.91579 & 140 & 3.5 & 3.691 & $21\pm19$  & $0.17\pm0.85$  & $15.2\pm0.3$ & $0.2\pm0.3$  & $22.7\pm0.3$ \\
2017 Jan 21 & 07:27:14 & 38.81418 & 160 & 3.0 & 4.235 & $19\pm14$  & $-2.08\pm0.30$ & $18.8\pm0.3$ & $-0.4\pm0.3$ & $22.2\pm0.3$ \\
2017 Jan 23 & 07:30:47 & 40.81649 & 160 & 3.1 & 4.457 & $33\pm17$  & $-0.72\pm0.34$ & $15.0\pm0.3$ & $2.1\pm0.3$  & $27.4\pm0.3$ \\
2017 Feb 04 & 06:42:27 & 52.78190 & 170 & 3.0 & 5.787 & $7\pm12$   & $-0.55\pm0.51$ & $16.4\pm0.3$ & $-0.1\pm0.3$ & $24.6\pm0.3$ \\
2017 Feb 05 & 06:26:56 & 53.77103 & 150 & 3.3 & 5.897 & $-19\pm12$ & $-1.91\pm0.70$ & $16.8\pm0.3$ & $0.2\pm0.3$  & $23.7\pm0.3$ \\
2017 Feb 14 & 07:33:52 & 62.81665 & 160 & 3.1 & 6.902 & $16\pm12$  & $-2.42\pm0.69$ & $16.5\pm0.3$ & $0.5\pm0.3$  & $23.8\pm0.3$ \\
2017 Feb 15 & 07:22:41 & 63.80880 & 160 & 3.0 & 7.012 & $85\pm13$  & $-2.38\pm0.66$ & $16.3\pm0.3$ & $0.6\pm0.3$  & $22.2\pm0.3$ \\
2017 Feb 16 & 07:27:28 & 64.81202 & 150 & 3.3 & 7.124 & $93\pm14$  & $-1.51\pm0.59$ & $17.7\pm0.3$ & $-0.1\pm0.3$ & $23.1\pm0.3$ \\
2017 Feb 17 & 07:37:19 & 65.81876 & 180 & 2.7 & 7.235 & $22\pm18$  & $-1.84\pm0.49$ & $19.0\pm0.3$ & $-0.3\pm0.3$ & $19.8\pm0.3$ \\
\hline
\end{tabular}
\label{tab:log}
\end{table*}

\section{Spectropolarimetric observations}
\label{sec:obs}

Our set of observations consists of 18 circularly polarized spectra 
collected over 2 months from mid December 2016 to mid February 2017 with the ESPaDOnS spectropolarimeter at CFHT, covering 
the domain 370--1,000~nm at a resolving power of 65,000 \citep{Donati03}.  Raw frames were reduced with the standard ESPaDOnS 
reduction package ({\tt Libre ESpRIT}), and Least-Squares Deconvolution \citep[LSD,][]{Donati97b} was applied to all reduced spectra, 
using a line list appropriate to CI~Tau.  The full journal of observations is presented in Table~\ref{tab:log}.  The first Stokes $I$ 
spectrum of CI~Tau we collected (at cycle 0.011) was slightly affected by moonlight whose spectrum shows up in the far blue wing of 
the corresponding LSD profile.  This pollution, occurring far enough in the profile wings not to affect our analysis, was nonetheless 
filtered with the technique devised in this purpose \citep{Donati17}.  

\begin{figure*}
\includegraphics[scale=0.44,angle=-90]{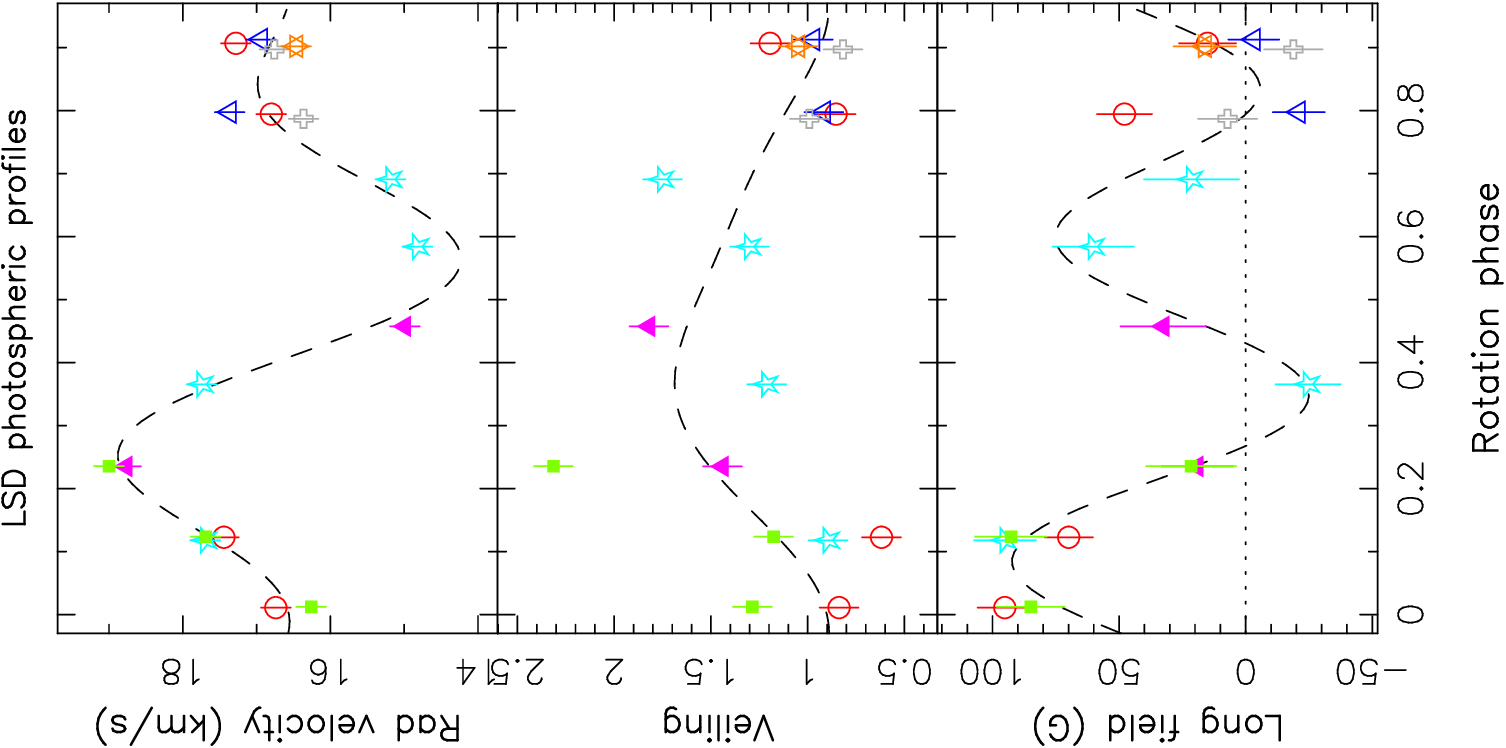}\hspace{3mm}
\includegraphics[scale=0.44,angle=-90]{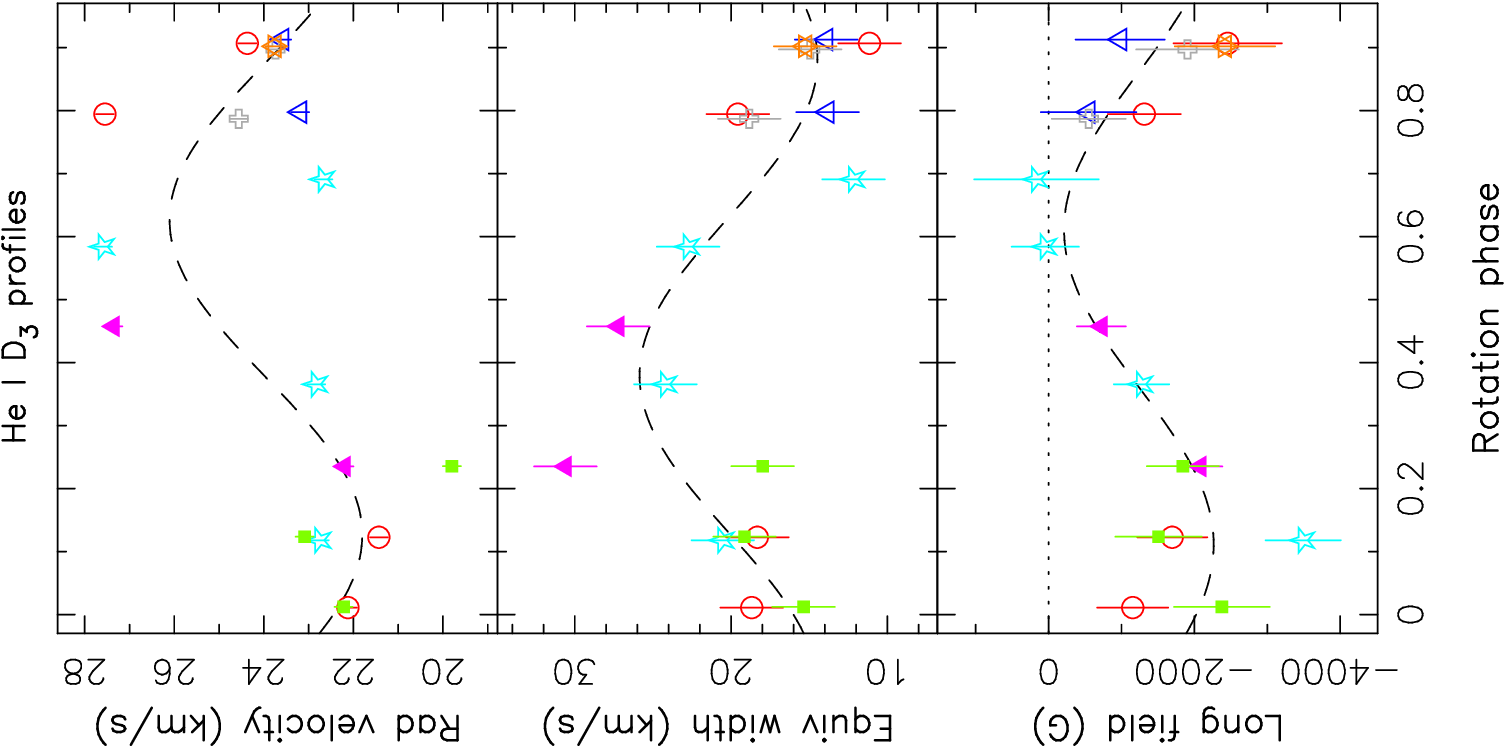}\hspace{3mm}
\includegraphics[scale=0.44,angle=-90]{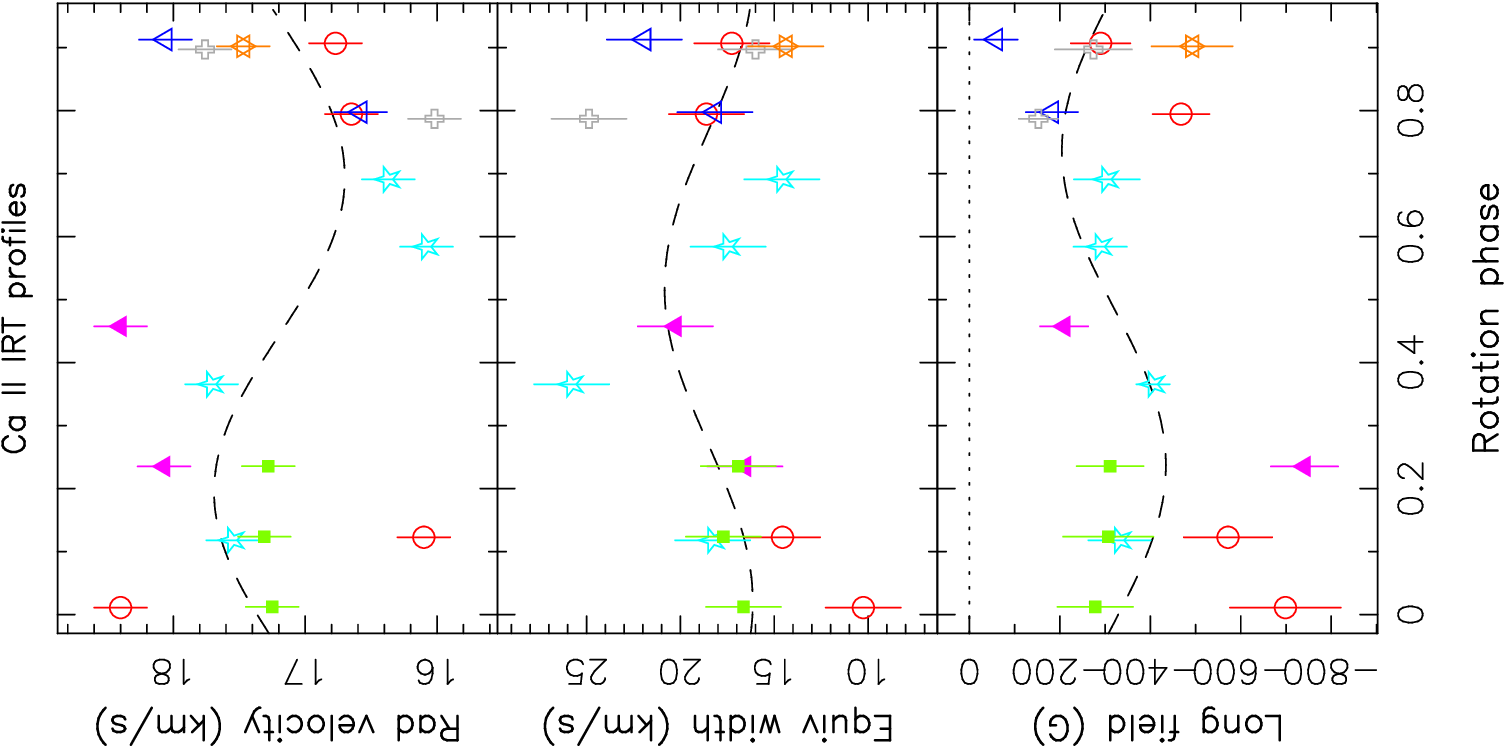} 
\caption[]{Variability and modulation of the LSD profiles (left panel), \hei\ $D_3$ (middle panel) and \caii\ IRT (right panel) 
NCs of CI~Tau as a function of rotation phase (computed with the ephemeris of Table~\ref{tab:log} assuming a 
rotation period of 9.0~d).  Each panel shows the RVs (top), the veiled EWs (or veiling in the case of LSD profiles, middle) and 
the longitudinal fields (bottom).  The red open circles, blue open triangles, cyan open pentagrams, purple filled triangles, 
open grey pluses, orange open hexagrams and filled green squares respectively depict measurements obtained during rotation 
cycles 0, 2, 3, 4, 5, 6 and 7.  The dashed lines show sine (plus first harmonic in the case of LSD profiles) fits to the data 
to emphasize modulation.  Positive longitudinal fields correspond to fields pointing towards the observer.  } 
\label{fig:var}
\end{figure*}

\begin{figure}
\center{\includegraphics[scale=0.27,angle=-90]{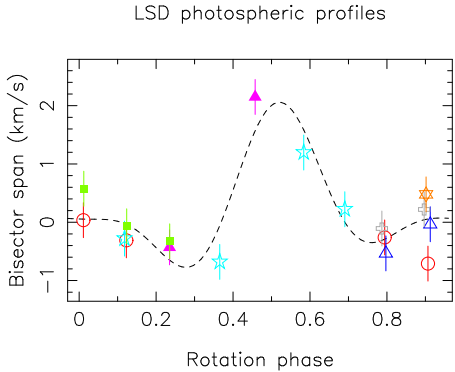}} 
\caption[]{Same as Fig.~\ref{fig:var} for the LSD-profile BSs of CI~Tau.  The sine fit to the data now includes the first two 
harmonics to better track the rapid temporal variability of BSs \citep[e.g.,][]{Hebrard14}}. 
\label{fig:bis}
\end{figure}

Radial velocities (RVs) of the Stokes $I$ LSD photospheric profiles of CI~Tau (computed from the first moment of the unpolarized 
profiles, see Fig.~\ref{fig:var} top left panel) are clearly modulated with a period of $9.02\pm0.06$~d;  although the RV variations 
are largely sinusoidal, adding the first harmonics significantly improves the fit, with the corresponding reduced chi-square (\chisqr) 
showing a single deep and narrow minimum within a wide range of periods (5 to 16~d) and featuring a false alarm probability 
of order 0.1\%.    We also note that LSD profiles are exhibiting strong asymmetries at times, with bisector spans (BSs) varying with time (see 
Fig.~\ref{fig:bis}) and being modulated with a period of $8.9\pm0.1$~d.  

Zeeman signatures from large-scale surface magnetic fields are detected (with a false alarm probability smaller than 0.1\%, and 
in most cases smaller than 0.001\%) in all circular polarisation (Stokes $V$) LSD profiles of CI~Tau.  The corresponding longitudinal 
fields (i.e., line-of-sight projected magnetic fields averaged over the visible hemisphere) probed by Stokes $V$ LSD profiles range 
from $-25$ to 95~G (see Fig.~\ref{fig:var} bottom left panel), and are found to vary with a period $9.04\pm0.08$~d, i.e., very close 
to that of the RVs variations of Stokes $I$ LSD profiles;  adding the first harmonics to the sinusoidal fit is essential given the 
two unequal minima and maxima that the measured longitudinal fields exhibit over one period.  

\section{Veiling and accretion}
\label{sec:acc}

The amount of veiling\footnote{Veiling is a consequence of accretion, in particular shocks occurring at the surface of the star and 
potential interactions between the magnetosphere and the inner accretion disc \citep[e.g.,][]{Beristain01}, that generate a 
mostly featureless continuum, stronger in the blue than in the red, that adds up on the stellar spectrum and reduces the depths, or 
'veils', the line profiles.  \color{black} The values we show here (see Fig.~\ref{fig:var}) represent the excess continuum flux relative 
to that from the stellar photosphere, at an average wavelength of about 620~nm. } that the LSD profiles are subject to (compared to those of 
a wTTS of similar spectral type) is found to 
vary between 0.5 and 2.5 (see Fig.~\ref{fig:var} middle left panel).  These fluctuations feature a significant amount of intrinsic 
variability, as is often the case with veiling and more generally with accretion-related phenomena;  they also include a modulated 
component exhibiting a period of $8.7\pm0.2$~d, slightly smaller though still compatible with that inferred from the modulation of RVs 
and longitudinal fields.  By computing LSD signatures for blue and red spectral regions separately (with weighted-average central 
wavelengths of 520 and 740~nm respectively), we find that both the average veiling and its modulation are 2.3$\times$ stronger in 
the blue than in the red, consistent with what is expected for veiling in cTTSs.  
\color{black} 
We also note that the veiling in our optical spectra 
of CI~Tau is typically 3$\times$ larger than that reported in previous studies \citep[e.g.,][]{Hartigan95, Beristain01, Herczeg14}, 
suggesting that the star is either more actively accreting and / or that its photosphere is intrinsically fainter (with respect to 
the main veiling sources).  

\begin{figure*}
\includegraphics[scale=0.6,angle=-90]{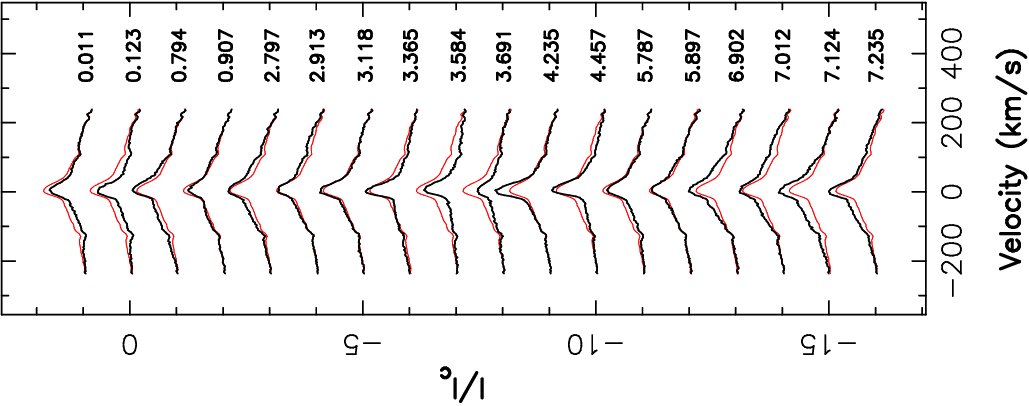}\hspace{2mm}
\includegraphics[scale=0.6,angle=-90]{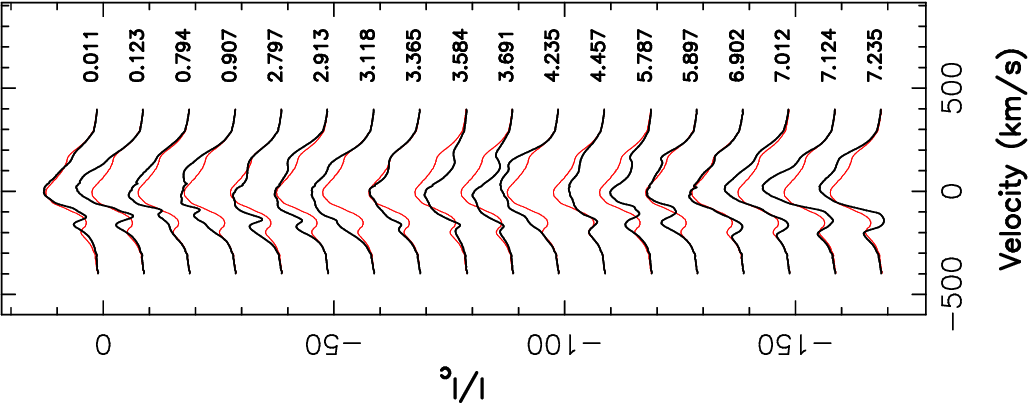}\hspace{2mm}
\includegraphics[scale=0.6,angle=-90]{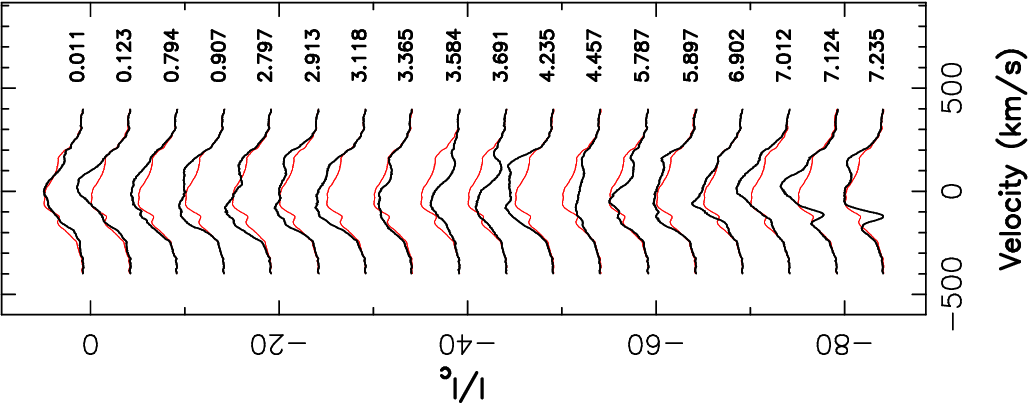}\hspace{2mm}
\includegraphics[scale=0.6,angle=-90]{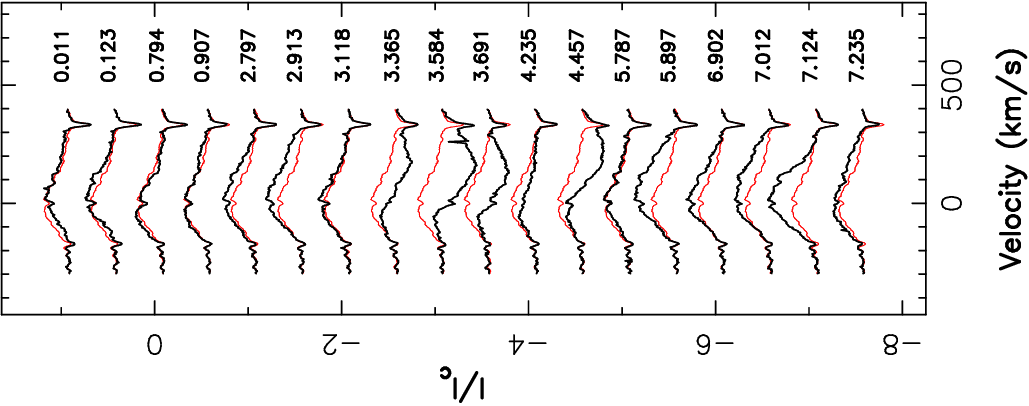}
\caption[]{From left to right, \color{black} \hei\ $D_3$, \color{black} \hal, \hbe\ and \color{black} 777.19~nm \oxi\ profiles profiles of CI~Tau 
\color{black} throughout our observing run.  The red line depicts the average profile over the run to emphasize temporal variations.  
\color{black} Note the different scales on both $x$ and $y$ axes. }
\label{fig:lin}
\end{figure*}

Accretion at the surface of the star is best probed with the \hei\ $D_3$ line complex at 587.56~nm (see Fig.~\ref{fig:lin} left panel), 
exhibiting in particular less intrinsic variability than veiling and offering at the same time the additional capability of estimating 
magnetic fields within accretion regions \citep{JohnsKrull99}.  In cTTSs actively accreting from their inner discs like CI~Tau, this 
\hei\ feature usually exhibits both a narrow core emission component (NC) probing the post-shock region close to the stellar surface      
and located at the footpoint of the magnetized funnels, and a broad emission component (BC) coming either from the accretion 
funnels linking the star to the inner disc and / or from a hot wind likely powered by interactions between the magnetosphere and the 
inner disc \citep{Beristain01}.  
In agreement with this picture, we find that Stokes $V$ Zeeman signatures in the \hei\ $D_3$ line of CI~Tau are only detected in 
conjunction with the NC, i.e., where the magnetic field is expected to be strongest and the velocity gradient of the emitting plasma 
to be smallest.  The BC of CI~Tau is typically 5$\times$ broader and stronger than the NC, which features an average (veiled) 
equivalent width (EW) and full-width-at-half-maximum of 20 and 40~\kms\ (0.04 and 0.08~nm) respectively when most intense.  

Both EWs and RVs of the \hei\ $D_3$ NC are clearly modulated with time in CI~Tau (see Fig.~\ref{fig:var} middle and top central 
panels), with respective periods of $9.1\pm0.2$ and $9.0\pm0.2$~d.  EWs and RVs of the BC (not shown) are also variable with time, 
but anti-correlated with the NC;  when corrected for the veiling, the BC no longer shows consistent periodic fluctuations of its EW 
but rather intrinsic variability about an average strength of about 240~\kms\ (0.47~nm), whereas the NC exhibits an even stronger 
modulation of its EW, now peaking at about 70~\kms\ (0.14~nm).   
Zeeman signatures are detected at most epochs in the Stokes $V$ profile of the \hei\ $D_3$ NC of CI~Tau, probing 
magnetic fluxes at the bottom of accretion funnels.  The corresponding longitudinal fields are found to range from 0 to $-2.5$~kG (and 
even $-3.5$~kG in a single case, see Fig.~\ref{fig:var} bottom central panel) with a median error bar of 0.5~kG, and to be modulated 
with a period of $8.9\pm0.2$~d, in good agreement with all other periods identified so far in our data set.  

The \caii\ infrared triplet (IRT) was also shown to probe accretion regions of cTTSs \citep[e.g.,][]{Muzerolle98}, albeit more  
ambiguously than the \hei\ $D_3$ line as a result of the chromospheric activity often contributing significantly to the NC 
\citep{Donati11}.  In the case of CI~Tau, the \caii\ IRT also features both a NC and a BC, with the BC being typically 10$\times$ 
broader and 40$\times$ stronger than the NC.  Given the strong intensity contrast between both components and the often 
complex shape of the BC, reliably extracting (with a multiple Gaussian fit) the much weaker NC is tricky.  Whereas the 
NC exhibits RVs that vary with a period of $9.1\pm0.6$~d, its EW, equal to 18~\kms\ or 0.05~nm in average (in the 
veiled spectrum), is not seen to fluctuate in a clear periodic way (see Fig.~\ref{fig:var} top and middle right panels), 
presumably as the result of a high level of intrinsic variability in the dominant contribution from chromospheric activity.  
Zeeman signatures in the Stokes $V$ LSD profiles of the \caii\ IRT NC are detected most of the times, with longitudinal 
fields in the range $-60$ to $-740$~G with a median error bar of 70~G (see Fig.~\ref{fig:var} bottom right panel).  These 
longitudinal fields are typically 5$\times$ weaker than those inferred from the \hei\ NC, indicating that accretion contributes 
to about 20\%\ of this emission whereas chromospheric activity contributes to 80\%\ (assuming that magnetic fields in accretion 
regions are much stronger than those in non-accreting ones, as expected from previous studies).  As for EWs, longitudinal fields 
of the \caii\ IRT NC exhibit a high level of intrinsic variability and show no more than a hint of periodicity (with a period 
of $8.0\pm0.3$~d).  

Given that most spectral features of CI~Tau, and in particular photospheric LSD profiles and the \hei\ $D_3$ NC reliably 
probing accretion post-shock regions near the surface of the star, are observed to fluctuate in a regular way over our 2-month 
observing run with a clear period of 9.0~d\footnote{The weighted average of all periods measured from both LSD profiles and the 
\hei\ $D_3$ NC is $9.00\pm0.05$~d.}, we conclude that this period is almost certainly the rotation period of the star.  
Moreover, the reported variability of CI~Tau is found to be very similar to that of other cTTSs \citep{Donati11, Donati11b, Donati12}, 
coming as further evidence that what we see is truly rotational modulation.  It logically follows that the main 9~d period detected 
in the photometric light curve of CI~Tau collected with the K2 space probe during campaign 13 on Taurus between 2017 March~07 and 
May~27 \citep{Biddle18}, i.e., about 1--3 months after the end of our ESPaDOnS run, is attributable to rotational modulation 
caused by brightness features at the surface of the star carried 
in and out of the observer's view.  We stress that the 9-d period is not only the dominant one, but also the only one that shows 
up consistently in all our time series;  by contrast, the period of 6.6~d also present in the K2 data (as the third strongest peak 
in the periodogram in the period range 5--16~d) and previously suggested to be the rotation period of CI~Tau \citep{Biddle18}, is 
not seen in the (veiled or unveiled) EW of the \hei\ NC (whose 9-d modulation is quite clear, see Fig.~\ref{fig:var} middle 
central panel), and only marginally (in the range 6.8--7.0~d, along with several other weak peaks of comparable strengths) in 
the RVs and longitudinal fields of the \hei\ line and LSD photospheric lines.  We thus phased our data using a rotational period 
of 9.0~d (see ephemeris in Table~\ref{tab:log}).  

In this context, our data (see Fig.~\ref{fig:var}) suggest that the main accretion region is located at phase 0.3--0.4, i.e., when 
both the \hei\ NC and the veiling are strongest.  Bisector variations in LSD profiles further confirm that the 9-d modulation is 
caused by the presence of surface features, known to distort spectral features, rather than by, e.g., an orbiting body that would 
shift the spectral lines without affecting their shapes and bisectors.  RV curves of LSD profiles and of \hei\ NC further indicate 
that the accretion region is dark at photospheric level and bright at chromospheric level, and that the post-shock zone in which the 
\hei\ NC forms is in average red-shifted with respect to the stellar rest-frame by about 7~\kms.  The main accretion region 
apparently coincides with strong magnetic fields of up to $\simeq$3~kG, causing it to appear dark at photospheric level with respect 
to the less-magnetic, non-accreting surroundings, and thereby making the strong fields to go undetected in LSD profiles.  We 
also report that the K2 brightness of CI~Tau is minimum (by about 15\%) when veiling is maximum, indicating that the observed 
brightness modulation mostly relates to the surface distribution of photospheric features and in particular to the presence of the 
dark accretion region mentioned above;  it even suggests that veiling variations are in fact dominated by the periodic changes in the 
photospheric brightness of CI~Tau rather than by fluctuations in the veiling sources themselves, consistent with the fact that the 
the intensity of the \hei\ BC (associated with the main veiling source) exhibits no obvious rotational modulation.  
More quantitative results require a full tomographic analysis, which we carry out in the following section.  

From the unveiled peak EW of the \hei\ NC that we observe for CI~Tau (about 70~\kms\ or 0.14~nm once averaged over the successive 
cycles), we can compute the peak logarithmic flux in the line ($-4.5$ in units of \lsun), translate it into a peak logarithmic 
accretion luminosity \citep[$-1.1$ in units of \lsun, using empirical calibrations of][]{Fang09}, and estimate a logarithmic 
mass accretion rate of $-8.2\pm0.3$ (in units of \mspy).  Repeating the same calculation for the (unmodulated) \hei\ BC, we derive 
logarithmic line flux, accretion luminosity and accretion rate of $-3.9$, $-0.4$ and $-7.4$ respectively (in the same units as for 
the NC).  At this point, it is unclear whether the mass accretion rates derived independently from the \hei\ NC and BC have any 
physical interpretation, or whether they only make sense when considered together;  we come back on this point in Sec.~\ref{sec:dis}.  
The contrast between the accretion luminosities of the \hei\ BC and NC (about 0.7 in log) is consistent with the ratios of 
featureless continua associated with both components, which can be estimated from the observed modulation amplitudes of the veiling 
and of the K2 photometric flux, and from the assumption that only the continuum associated with the NC happens to be modulated 
with rotation (in phase with the EW of the NC).  

Balmer lines of CI~Tau (see Fig.~\ref{fig:lin} middle panels), also in emission, feature average (unveiled) equivalent widths of 8500 
and 3800~\kms\ respectively (18.6 and 6.2~nm), translating into logarithmic line fluxes, accretion luminosities and mass accretion rates 
of $-2.3$, $-0.6$ and $-7.7$ for \hal\ and $-2.9$, $-0.7$ and $-7.8$ for \hbe\ (in units of \lsun, \lsun\ and \mspy).  

Both lines 
also feature conspicuous red-shifted absorption components (with respect to the mean profile) at cycles 3.584, 3.691, 4.457 and 5.787, 
likely tracing accretion funnels transiting the stellar disc as the star rotates, suggesting that accretion occurs towards polar (rather 
than equatorial) regions given the inclination of the rotation axis (see Sec.~\ref{sec:evo}).  
Even stronger red-shifted absorption is observed in the 777.19~nm 
\oxi\ line (see Fig.~\ref{fig:lin} right panel) with the red wing extending up to a velocity of 380~\kms\ with respect to the star at 
cycles 3.584 and 4.457, corresponding to $92\pm7$\%\ of the escape velocity of CI~Tau (equal to $\vesc=415\pm30$~\kms, see 
Table~\ref{tab:par}).  

Finally we report time-variable blue-shifted narrow absorption features in both \hal\ and \hbe\ at velocities in excess of 100~\kms\ 
(see Fig.~\ref{fig:lin} middle panels, at cycles 7.124 and 7.235, but also present in \hal\ only at many other phases), suggesting 
that intermittent magnetospheric ejection phenomena are taking place in the circumstellar environment of CI~Tau.  
We come back on these points in Sec.~\ref{sec:dis}.  
\color{black}

\section{Tomographic modelling of CI~Tau}
\label{sec:tom}

To recover the large-scale magnetic field at the surface of CI~Tau simultaneously with maps of the photospheric brightness and of the 
accretion-induced emission in the \hei\ $D_3$ line at chromospheric level, we use the tomographic approach called Zeeman-Doppler Imaging.  
This is achieved as in previous similar papers \citep[e.g.,][]{Donati19}, with one significant difference.  
In the present study, we used the \hei\ $D_3$ NC as the accretion proxy, rather that the \caii\ IRT profiles that were 
found to be poorly informative about accretion regions at the surface of CI~Tau (see Sec.~\ref{sec:acc}).  The main disadvantage of 
the \hei\ $D_3$ NC is that they usually exhibit asymmetric local Stokes $I$ profiles, and local Stokes $V$ profiles departing significantly 
from antisymmetry \citep[i.e., with stronger / narrower blue lobes and shallower / broader red lobes on both sides of the line center, see, e.g., 
the top panel of Fig~3 in][as a result of the velocity gradients in the line formation region]{Donati11}, rendering them less straightforward 
to model than the more standard \caii\ IRT profiles.  In the case of CI~Tau however, the intrinsic variability of the \caii\ IRT NC
gives us no other choice than to apply our model to the \hei\ $D_3$ profiles.  

The main steps of the imaging process are as follows.  Starting from featureless images, ZDI iteratively adds information on the stellar 
surface maps to optimise the fit to the observations using conjugate gradient techniques, until data are fitted at a given \chisqr\ level.  
Maps of the photospheric brightness (assuming cool spots only) and of the accretion-induced \hei\ NC (assuming bright 
chromospheric accretion regions only) are described as grids of 
independent pixels, whereas the poloidal and toroidal components of the large-scale fields are expressed as series of spherical harmonics 
\citep[SH,][]{Donati06b}, truncated to modes with $\ell\leq7$ in the case of CI~Tau.  We further assume that the field is mostly 
antisymmetric with respect to the centre of the star \citep[i.e., that odd SH modes dominate, as in, e.g.,][]{Donati11}, so that 
accretion funnels linking the inner disc to the star are anchored at high latitudes, as often reported for cTTSs (e.g., from the 
amplitude of the RV modulation of accretion lines, see Fig.~\ref{fig:var} top central panel).  

To approximate the local Stokes $I$ and $V$ profiles of both photospheric absorption and \hei\ NC profiles at each point of 
the stellar surface, we use Unno-Rachkovsky's analytical solution to the polarized radiative transfer equations in a Milne-Eddington
atmosphere, taking into account the local values of the various relevant parameters, i.e., radial velocity and limb darkening (for all 
maps), relative photospheric brightness (for the photospheric brightness and magnetic maps), accretion-induced \hei\ emission at 
chromospheric level (for the \hei\ emission and magnetic maps) and magnetic field (for all maps).  We then integrate local profiles 
over the whole visible hemisphere to infer the synthetic profiles of CI~Tau at each observed epoch.  A magnetic filling factor of 
30\% over the whole stellar surface was assumed for our study, applying both to LSD profiles and \hei\ NC as in our 
previous papers \citep[e.g.,][]{Donati10b}. Velocity gradients within the \hei\ line formation region are not taken into account, 
implying that our model is not capable of reproducing asymmetries of the local \hei\ Stokes $I$ and $V$ profiles.  

\begin{figure*}
\includegraphics[scale=0.6,angle=-90]{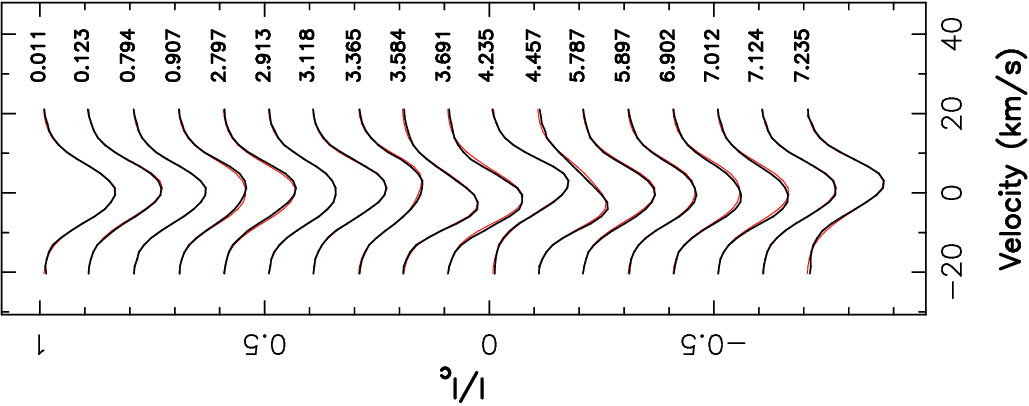}\hspace{2mm}
\includegraphics[scale=0.6,angle=-90]{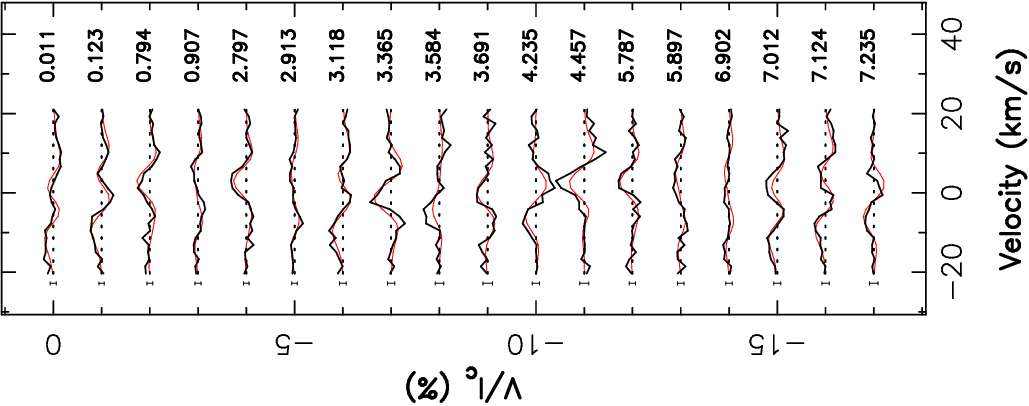}\hspace{3mm}
\includegraphics[scale=0.6,angle=-90]{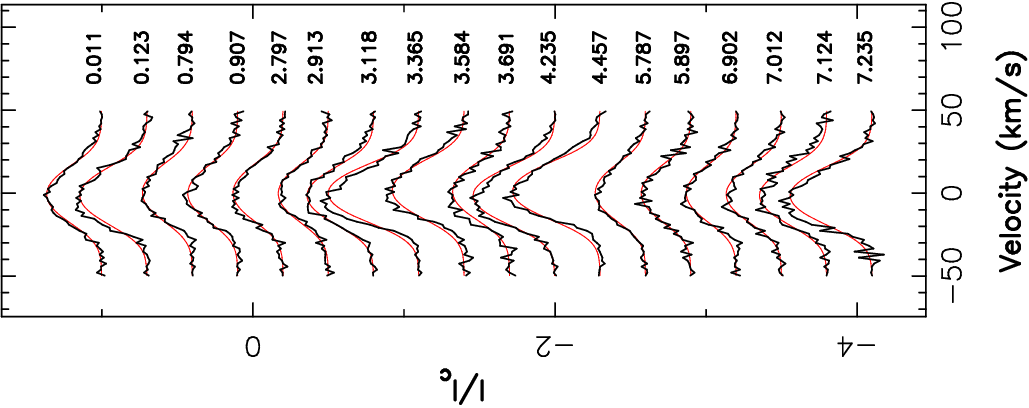}\hspace{2mm}
\includegraphics[scale=0.6,angle=-90]{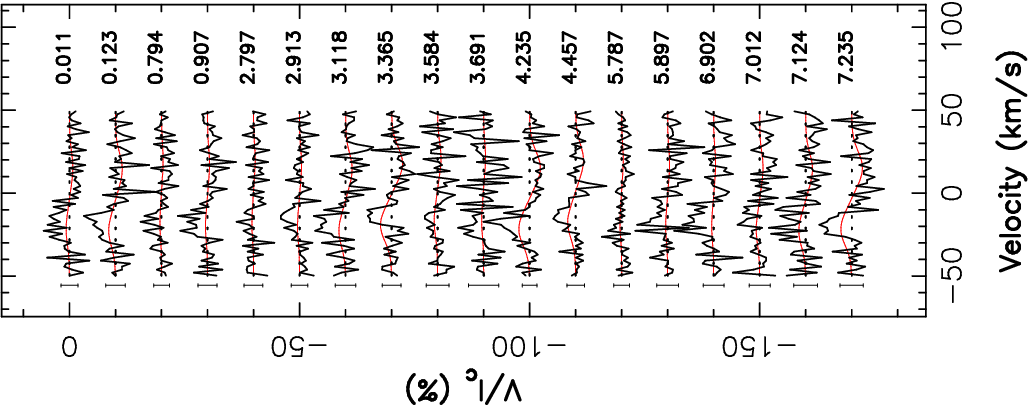}
\caption[]{Observed (thick black line) and modelled (thin red line) LSD Stokes $I$ and $V$ profiles of the photospheric lines 
(left panels) and of the \hei\ $D_3$ core-emission profiles (right panels) of CI~Tau.  Rotation cycles and $\pm$1$\sigma$ error bars 
(for Stokes $V$ profiles only) are indicated to the right and left of each observation respectively.} 
\label{fig:fit}
\end{figure*}

The fit to our data that ZDI obtains is shown in Fig.~\ref{fig:fit} for both the LSD photospheric and \hei\ NC profiles 
(left and right panels respectively).  Overall, we find that Stokes $I$ and $V$ profiles and their first order moments (i.e., the RVs 
and longitudinal fields) are well reproduced, especially given the large amount of intrinsic variability that the spectra of CI~Tau 
exhibits.  As expected from the limitations of our model, the fit to the Stokes $V$ profiles of the \hei\ core emission is less 
accurate (see below for a more detailed description).  We find that the \vsini\ providing the best match to the far wings of the 
Stokes $I$ profiles is $9.5\pm0.5$~\kms, whereas the optimal average RV of CI~Tau is $\vrad=16.8\pm0.2$~\kms\ 
\citep[consistent with most young stars in the Taurus cloud,][]{Galli18}.  

One can note from Fig.~\ref{fig:fit} (left panel) that the RV variations of LSD profiles (reaching a peak-to-peak amplitude of 4~\kms, 
see Fig.~\ref{fig:var}) and the corresponding BS changes (see Fig.~\ref{fig:bis}) are attributable to profile distortions and 
asymmetries (e.g., at cycles 3.584 and 4.457) rather that to global line shifts (caused by, e.g., an orbiting body) that would generate neither 
profile asymmetries nor bisector changes;  this demonstrates that the RV curve of CI~Tau reflects mostly stellar activity and more 
specifically the presence of surface features.  We also note that the amplitude of the RV modulation is found to be slightly larger 
in the blue than in the red (by about 10\% between 520 and 740~nm), as expected from activity \citep[e.g.,][]{Reiners10}.  

\begin{figure*}
\includegraphics[scale=0.7,angle=-90,bb=35 40 434 709]{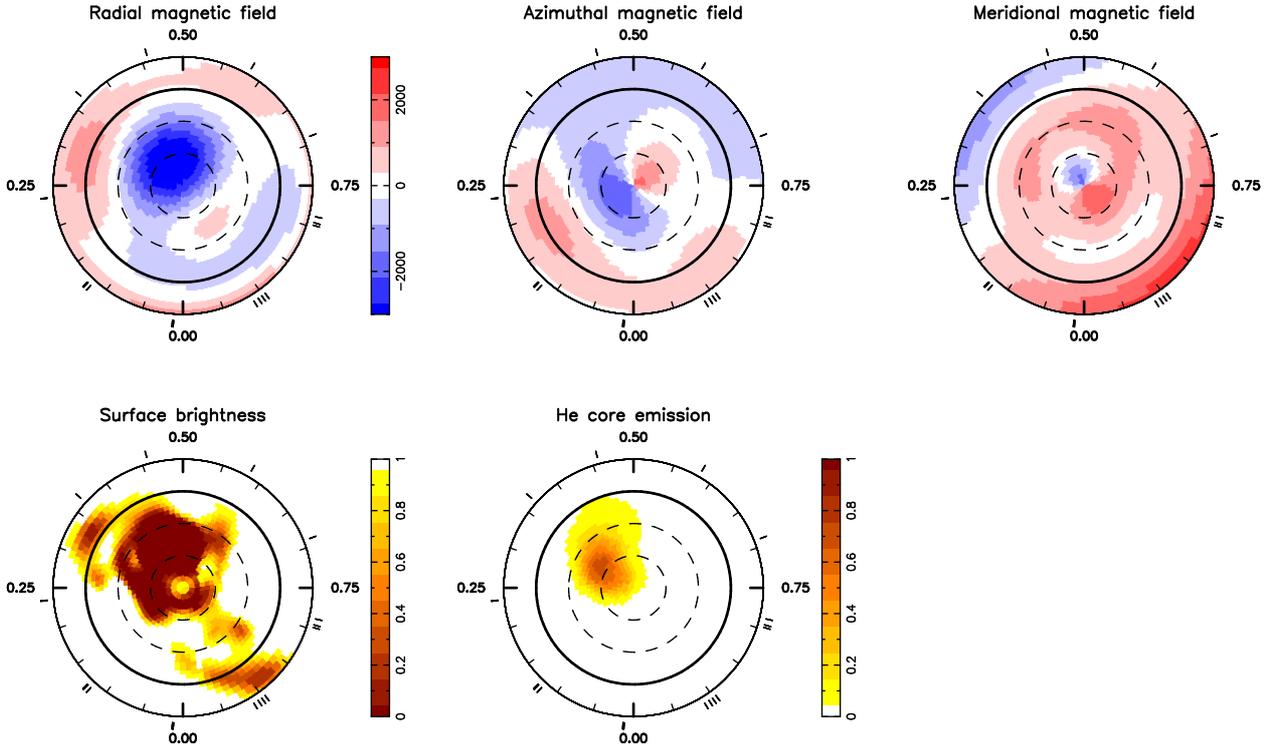}
\caption[]{Reconstructed maps of the magnetic field (top left, middle and right panels for the radial, azimuthal and 
meridional components in spherical coordinates, all in G), relative photospheric brightness (bottom left) and 
accretion-induced \hei\ emission in the NC (bottom  right)  at  the  surface  of  CI~Tau, derived from the data set of 
Fig.~\ref{fig:fit} using tomographic imaging.  The star is shown in a flattened polar projection down to a latitude of 
$-30$\degr, with the north pole at the center and the equator shown as a bold line.  Outer ticks depict phases of observations.
Positive radial, azimuthal and meridional fields respectively point outwards, counterclockwise and polewards. } 
\label{fig:map}
\end{figure*}

This conclusion is further confirmed by ZDI, 
demonstrating that CI~Tau hosts a high-latitude dark spot at phase 0.35--0.40, large enough (about 20\% of the overall stellar surface, 
see Fig.~\ref{fig:map}, bottom left panel) to generate the observed variability of LSD profiles and the corresponding RV fluctuations 
(with the change from maximum to minimum RV occurring from phase 0.25 to 0.55, see Fig.~\ref{fig:var}).  
Although the main reconstructed dark spot itself is quite reliable, its detailed shape and the surrounding low-latitude features and 
appendages are subject to caution as they may reflect tomographic imaging spuriously translating stochastic profile variability 
(caused by accretion) into small surface structures with limited visibility, when phase coverage is moderate.  

The surface map of \hei\ emission in the NC clearly shows a well-defined accretion region located at phase 0.35 and latitude 60\degr, and 
covering $\simeq$3\% of the overall stellar surface (see Fig.~\ref{fig:map}, bottom right panel).  This is again consistent with the 
phase at which emission in the \hei\ NC peaks, as well as with the observed RVs of the \hei\ NC (changing from 
from minimum to maximum RV between phase 0.1 and 0.6, see Fig.~\ref{fig:var}).  ZDI performs a fair job at reproducing the 
changes in the NC profile (see Fig.~\ref{fig:fit} third panel), although not the sharp blue side and smooth red side at maximum 
emission (e.g., at cycle 3.365) that reflect the intrinsic asymmetry of the local \hei\ NC which our model does not incorporate.  

The large-scale magnetic field that ZDI reconstructs is mostly poloidal, with a toroidal component that encompasses only $\simeq$15\% of the 
reconstructed magnetic energy (see Fig.~\ref{fig:map}, top panels).  Its main feature is a large (negative) radial field region at phase 0.35 and 
latitude 60\degr, where the field flux reaches values down to $-3.7$~kG, in agreement with peak longitudinal fields derived from \hei\ emission 
profiles;  this feature roughly overlaps with the extended dark spot at photospheric level and the main accretion region at chromospheric level 
that our Stokes $I$ LSD and \hei\ profiles revealed, confirming that accretion at the surface of CI~Tau takes place at its high-latitude 
magnetic poles.  The reconstructed toroidal field component causes magnetic field lines in the main radial field region to be more tilted 
towards the equatorial plane than those of the poloidal component;  this is likely a real feature and a direct consequence of 
on-going accretion from the disc.  

We note that the shapes of LSD Stokes $V$ profiles are well fitted by our model (see Fig.~\ref{fig:fit} second panel), with small departures 
from the data (e.g., at cycle 4.457) presumably attributable to intrinsic variability.  
The shapes of Stokes $V$ \hei\ NC profiles, and in particular the stronger / narrower blue lobes and the shallower / broader red lobes of the 
detected Zeeman signatures, are less well reproduced, with only their average intensities being matched by ZDI;  this again reflects the 
limitation of our model at describing local Stokes $V$ profiles at the surface of the star that are not antisymmetric with respect to the 
local line center.  This limitation is however not critical as our simple model is still able to grasp the fields needed to reproduce the 
average sizes of the blue and red lobes of the detected Zeeman signatures.  

The inferred large-scale field of CI~Tau mainly consists of a $-1.7$~kG dipole component (tilted at 20\degr\ with respect to the rotation axis 
towards phase 0.5 and enclosing about 60\% of the reconstructed magnetic energy) and of an octupole component of similar strength, with both 
being more or less aligned and parallel to each other.  Depending on how the respective weights of Stokes $I$ and $V$ profiles are balanced 
in the imaging process, we infer dipole strengths varying by about 15\%, i.e., from 1.5 to 2.0~kG (corresponding to 40--80\% of the 
reconstructed magnetic energy).  

We stress that modelling both LSD and \hei\ emission profiles simultaneously is essential to reliably unveil the main properties of the 
large-scale field of CI~Tau.  For instance, the dipole fields ZDI infers when applied to \hei\ or LSD profiles only are respectively equal 
to 3.9~kG and 0.3~kG, i.e., over- or under-estimated by typical factors ranging from 2.3$\times$ to 6$\times$ with respect to the value 
inferred from modeling both sets consistently;  
\color{black}
whereas Zeeman signatures of the \hei\ NC are critical to constrain field strengths within accretion regions \citep[as demonstrated 
in][]{Donati07}, simultaneously adjusting photospheric LSD profiles ensures that the relative areas of accreting and non-accreting 
regions, that depends on the relative intensities of the dipole and octupole components, are consistent with observations.  
\color{black}

\section{Summary and discussion}
\label{sec:dis}

Using ESPaDOnS at CFHT, we carried out a spectropolarimetric monitoring of the cTTS CI~Tau from mid December 2016 to mid February 
2017.  From a detailed analysis of its spectral lines, we find that CI~Tau is a $\simeq$2~Myr star with a mass of $0.9\pm0.1$~\msun\ 
and a radius of $2.0\pm0.3$~\rsun\ according to the evolution models of \citet{Siess00};  the mass we derive agrees well with the 
most recent dynamic estimate from radio interferometry \citep[$0.90\pm0.02$~\msun,][]{Simon19}.  Clear Zeeman signatures are detected in 
both photospheric and accretion lines of CI~Tau, demonstrating that the star hosts a strong large-scale magnetic field.  

Stokes $I$ and $V$ LSD profiles of photospheric lines and of the \hei\ $D_3$ accretion-induced NCs are found to be modulated 
with a clear period of $9.00\pm0.05$~d, with temporal variations (e.g., in RVs and BSs) that are similar to what is observed on 
other cTTSs so far \citep[e.g.,][]{Donati11, Donati19}, and phenomenologically consistent with what is expected from surface features 
being carried in and out of the observer's view by rotation \citep[e.g.,][]{Hebrard14}.  We thus conclude that the observed 9-d 
modulation is unambiguously attributable to the rotation period of the star, and so are the photometric fluctuations at the same period 
that K2 revealed through continuous observations collected about 1--3 months after the end of our ESPaDOnS run \citep{Biddle18}.  

Using ZDI, we successfully modelled the parent large-scale magnetic field of CI~Tau from the phase-resolved Zeeman signatures of both 
LSD photospheric and \hei\ NC profiles.  We find that CI~Tau possesses a strong, mainly poloidal field, featuring a dipole 
component of about $-1.7$~kG tilted at 20\degr\ with respect to the rotation axis, as well as a nearly aligned and parallel octupole field 
of similar strength, causing the radial field of CI~Tau to peak at $-3.7$~kG in a high-latitude region at phase 0.35.  
By simultaneously modelling the relative photospheric brightness 
and the accretion-induced NC of the \hei\ $D_3$ line, we show that the strong radial field region of CI~Tau overlaps with a 
large dark spot at photospheric level and with a bright accretion region at chromospheric level, demonstrating that this area at the 
surface of the star is the main footpoint of magnetic accretion funnels linking the inner disc to the star.  We note that the phase at 
which the dark spot is reconstructed in our photospheric brightness image (i.e., 0.35--0.40) agrees with that at which CI~Tau is 
statistically faintest as measured from the contemporaneous K2 photometry (once phased with our ephemeris, see Table~\ref{tab:log}), 
bringing still further support to the overall consistency of our tomographic modelling.  

\begin{figure}
\includegraphics[scale=0.36,angle=-90]{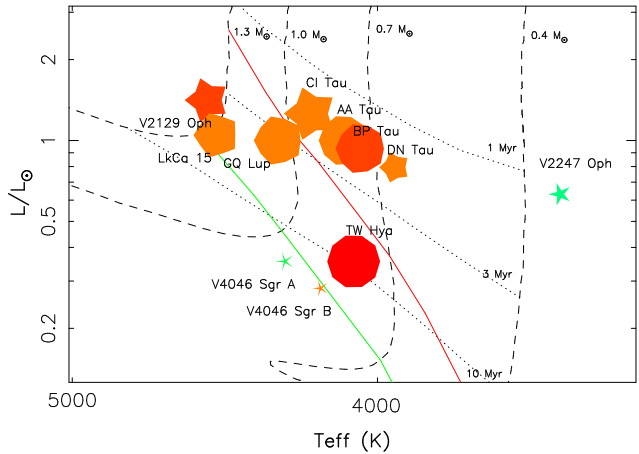}
\caption[]{Main properties of the large-scale magnetic fields of cTTSs, as a function of their position in the HR diagram. 
Symbol size indicates relative magnetic fluxes, symbol color illustrates field configurations (red to blue for purely poloidal 
to purely toroidal fields), and symbol shape depicts the degree of axisymmetry of the poloidal field component (decagon and stars for 
purely axisymmetric and purely non-axisymmetric poloidal fields, respectively).  The PMS evolutionary tracks and corresponding 
isochrones \citep{Siess00} assume solar metallicity and include convective overshooting. The full lines depict where models predict 
cTTSs start developing their radiative core (red line) and when their convective envelope gets shallower than 0.5~\rstar\ (green 
line) as they contract towards the main sequence. } 
\label{fig:tts}
\end{figure}

The magnetic field we reconstruct for CI~Tau is consistent with the evolutionary stage and internal structure of the host star, i.e., 
intense and dominantly poloidal and axisymmetric as for all fully- and largely-convective cTTSs magnetically imaged to date \citep{Gregory12}, 
and for all mature stars with masses in the range 0.25--0.50~\msun\ \citep{Morin08b, Morin10}.  We graphically summarize in 
Fig.~\ref{fig:tts} the magnetic properties of all cTTSs for which images are published in the literature, illustrating that all cTTSs 
more massive than 0.4--0.5~\msun\ and whose radiative envelope is thicker than about 0.5~\rstar\ are apparently able to trigger 
strong poloidal and mainly axisymmetric fields, whereas the fully-convective ones of them tend to have most of the poloidal magnetic 
energy concentrating into the dipole field component.  

\color{black} 
Depending on the assumed value of the logarithmic mass accretion rate (with potential estimates ranging from $-8.2$ for the \hei\ NC, 
to $-7.8$ and $-7.7$ for \hbe\ and \hal, to $-7.4$ for the \hei\ BC, all in units of \mspy, see Sec.~\ref{sec:acc}), we infer that 
the radius \rmag\ of the magnetospheric cavity that CI~Tau is able to carve at the centre of its accretion disc potentially ranges from 
6.3~\rstar\ down to 3.7~\rstar\ \citep{Bessolaz08}, or equivalently from 0.72 to 0.42~\rcor.  Taking the average value from the 
(unveiled) emission fluxes of \hal, \hbe\ and the \hei\ BC (yielding a logarithmic mass accretion rate of $-7.6\pm0.2$ in units of 
\mspy) as the most consistent estimate, we find that \rmag\ barely extends half way to \rcor.  
In this context however, it is hard to understand how 85\% of the accretion luminosity (i.e., the relative accretion luminosity 
associated with the \hei\ BC with respect to that of both \hei\ BC and NC) can be released before the accreted plasma even reaches 
the surface of the star;  if the accretion luminosity that fuels the \hei\ BC was indeed mostly released at the interface between 
the magnetosphere and the disc as suggested by \citet{Beristain01} for stars like CI~Tau, it should hardly exceed a small fraction 
of that associated with the \hei\ NC for magnetospheric radii of a few \rstar, since at such distances, the kinetic energy stored in 
the Keplerian motion is much smaller than the potential energy that can be released through free fall (e.g., by factors of 6 
and 14 at distances of 4 and 8~\rstar\ respectively).  
Moreover, it is 
unclear whether poleward accretion could occur (as observed) from within 0.5~\rcor\ given that the $\ell=1$ component of the field in 
the equatorial plane may not dominate the $\ell=3$ one by a large enough factor to efficiently deflect the accretion flow towards high 
latitudes at the stellar surface.  Finally, we would unlikely witness ejection events in the form of narrow blue-shifted emission 
features like those we observe in Balmer lines (see Sec.~\ref{sec:acc} and Fig.~\ref{fig:lin} middle panels) if the inner disc 
penetrates so much within \rcor.  We also note that accretion flows from 0.5~\rcor\ would imply free-fall velocities of 
0.87~\vesc, smaller (though still marginally compatible) with velocities observed in the red wing of the 777.19~nm \oxi\ line reaching 
up to 380~\kms\ (i.e., 0.92~\vesc, see Sec.~\ref{sec:acc} and Fig.~\ref{fig:lin} right panel) 

On the other hand, \rmag\ reaching out to 0.8--1.0~\rcor\ for an average logarithmic mass accretion rate of $-7.6\pm0.2$ (in units 
of \mspy) would require the dipole component of the large-scale field to be 2.4--3.6$\times$ stronger than what we infer (i.e., 
4.1--6.1~kG), which is not compatible with our spectropolarimetric data.  The logarithmic mass accretion rate derived from the \hei\ 
NC component only, equal to $-8.2$ (in units of \mspy) and bringing \rmag\ within 30\%\ of \rcor, i.e., in better agreement with 
some of the observed properties, is however most likely an underestimate in stars like CI~Tau where the \hei\ BC emission component 
dominates the NC, as pointed out by \citet{Beristain01}.  We suspect that the actual mass accretion rate of CI~Tau probably lies 
between the estimates inferred from the EWs of the \hei\ NC and BC, with the latter being likely sensitive to a wider hot plasma 
environment than just the disc / magnetosphere interface and the accretion funnels in stars featuring intense accretion and dominant 
BC for most accretion proxies.  

We stress again that our observations of CI~Tau were collected at a time where accretion was unusually strong, i.e., with accretion 
luminosities derived from the \hei\ BC and NC that are typically 5$\times$ larger than those inferred from \citet{Beristain01}.  
(In \citealt{Beristain01}, the unveiled \hei\ BC component of CI~Tau is about as strong as the unveiled NC component in our data).  
Assuming the observations of \citet{Beristain01} 
better represent the typical accretion status of CI~Tau would imply that \rmag\ usually falls within 0.7 and 1.0~\rcor\ (for accretion 
rates inferred from the EWs of the \hei\ BC and NC respectively), i.e., close enough to \rcor\ for CI~Tau to be in a situation of 
efficient spin-down through a propeller-like mechanism \citep{Romanova04, Ustyugova06, Zanni13} and in agreement with its 
longer-than-average rotation period (of 9.0~d).  

We also report that accretion on CI~Tau is highly unsteady, with the (unveiled) EW of the \hei\ BC (not modulated by rotation, see 
Sec.~\ref{sec:acc}) fluctuating by as much as a factor of 2 in a stochastic way on a timescale of only a few days.  
This erratic behaviour also shows up through the rapid photometric changes observed during the continuous 80-d monitoring campaign 
of K2, even though rotational modulation at the 9-d period is still clearly visible in the light curve.  
This intrinsic variability likely reflects a very inhomogeneous density profile in the inner regions of the accretion disc.  

More spectropolarimetric observations, in particular through campaigns monitoring the rotational modulation and stochastic variability 
of accretion proxies and photospheric lines at multiple epochs, are needed to constrain further the accretion geometry and 
magnetospheric properties of CI~Tau, now that the rotation period of the star is well determined.  
\color{black} 

The 9-d modulation of photospheric lines, both in position and shape (as inferred from RVs and BSs, see Figs.~\ref{fig:var} and 
\ref{fig:bis}), that results from the presence of brightness and magnetic features at the surface of CI~Tau (in particular the large dark 
spot and magnetic region at phase 0.35), generates in our optical spectra RV fluctuations with a semi-amplitude of $\simeq$2~\kms\ 
that are reminiscent of those reported by \citet{JohnsKrull16}, suggesting that they likely share the same origin.  The obvious 
distortions in the shape of LSD profiles (see Fig.~\ref{fig:fit}) and associated BS variations (see Fig.~\ref{fig:bis}) clearly 
indicate that the RV modulation we see on CI~Tau is attributable to the activity jitter\footnote{The small reduction in the amplitude of 
the RV modulation between the blue and red regions of our spectra further supports this interpretation.}, which is likely to show up at infrared 
wavelengths as well (with a different amplitude) given the strong magnetic fields detected at the surface of the star \citep{Reiners13}.  

Our result does not necessarily imply that the candidate close-in massive planet reported to orbit CI~Tau \citep{JohnsKrull16} does not exist;  
in fact, given the apparently very active planet formation going on in the protoplanetary disc of this young star \citep{Clarke18}, a hot 
Jupiter may indeed be present, orbiting at the outer boundary of the magnetospheric gap that almost extends out to the corotation radius.  
However, firmly demonstrating the existence of this putative hot Jupiter calls for definite evidence that is currently lacking, and that 
will likely be tricky to secure if the orbital period of the planet is indeed close to the rotation period of the star.  

\color{black} 
Recently, a CO signature, whose spectral location was reported to be modulated with the 9-d period and a semi-amplitude of 77~\kms, was 
detected in the spectrum of CI~Tau and putatively attributed to the candidate planet's atmosphere \citep{Flagg19}.  The RV modulation of 
this CO signature points to a structure located at a distance of \rcor\ from the centre of the star (see Sec.~\ref{sec:evo}), and crossing 
the line of sight when the dark photospheric spot is best visible (i.e., phase 0.4, see Sec.~\ref{sec:tom});  an alternate option to 
tentatively account for the observed spectral signature is thus that CO is present in the inner disc regions near \rcor, with an azimuthal 
structure linked to the accretion funnels and reflecting the star-disc magnetic coupling.  If confirmed, this would come as further evidence 
that the magnetosphere of CI~Tau is at times able to extend as far as \rcor.  Repeating such monitoring in spectropolarimetric mode, e.g., 
with an instrument like SPIRou \citep[][]{Donati17b}, now operational at CFHT, will allow one to quantify how this CO signature relates 
to the magnetosphere and to the large-scale field of the central star.  More generally, surveying the modulation of CO lines in cTTSs 
may come up as a novel way to independently estimate \rmag\ (by directly measuring the distance at which accretion funnels merge with 
the inner disc), and thereby to improve our description and understanding of magnetospheric accretion processes in young forming stars.  

Our study shows that CI~Tau is an ideal laboratory to study the detailed physics of star-disc interactions in low-mass stars, in particular 
those hosting magnetic fields strong enough to be able to trigger, at least part of the time, ejection processes capable of efficiently 
spinning down the star through a propeller mechanism.  This formation stage is indeed key for deciphering the early rotation history of 
Sun-like stars in the first few Myr of their life during which they vigorously interact with their discs \citep[e.g.,][]{Gallet19}.  
That CI~Tau is at the same time in a phase of active planet formation makes it even more interesting, for instance to find out whether it 
indeed hosts a massive planet that migrated close to the star after being formed further out in the disc.  Multi-wavelength campaigns 
involving simultaneously optical and near-infrared high-resolution spectropolarimeters like SPIRou, capable of revealing through 
imaging techniques what is occurring at the surface of the star, coupled to continuous precision photometry from space as obtained by K2 
or TESS, is the key to such studies aimed at unveiling the origins of worlds like or unlike our Solar System.  
\color{black}

\section*{Acknowledgements}
We thank an anonymous referee for valuable comments that enabled to clarify the analysis presented in this paper.  
Our study is based on data obtained at the CFHT, operated by the CNRC (Canada), INSU/CNRS (France) and the University of Hawaii.  
This project received funding from the European Research Council (ERC) under the H2020 research \& innovation programme (grant agreements 
\#740651 NewWorlds and \#742095 SPIDI).  SHPA acknowledges financial support from CNPq, CAPES and Fapemig.  
We also thank the Programme National de Physique Stellaire (PNPS) of CNRS/INSU for financial support.  
FM acknowledges funding from ANR of France under contract number ANR-16-CE31-0013.

\bibliography{citau}
\bibliographystyle{mnras}

\bsp	
\label{lastpage}
\end{document}